\documentclass[preprint,superscriptaddress,showpacs,amsmath,amssymb]{revtex4}
\usepackage{amssymb}
\usepackage{graphicx}
\usepackage{dcolumn}

\usepackage{color}

\begin{document}
\title{Quantum Phase transitional patterns in the SD-pair shell model}

\author{Yanan Luo}
\email{luoya@nankai.edu.cn} \affiliation{Department of Physics,
Nankai University, Tianjin, 300071, P.R. China}

\author{Yu Zhang}
\affiliation{Department of Physics and State key Laboratory of
Nuclear Physics and Technology, Peking University, Beijing, 100871,
P.R. China}

\author{Xiangfei Meng}
\affiliation{Department of Physics, Nankai University, Tianjin,
300071, P.R. China}

\author{Feng Pan}
\affiliation{Department of Physics, Liaoning Normal University,
      Dalian 116029, P. R. China}
\affiliation{Department of Physics and Astronomy, Louisiana State
University, Baton Rouge, LA 70803, USA}

\author{Jerry P. Draayer}
\affiliation{Department of Physics and Astronomy, Louisiana State
University, Baton Rouge, LA 70803, USA}

\date{\today}

\begin{abstract}
Patterns of shape-phase transition in the proton-neutron coupled
systems are studied within the $SD$-pair shell model. The results
show that some transitional patterns in the $SD$-pair shell model
are similar to the $U(5)-SU(3)$, $U(5)-SO(6)$ transitions with
signatures of the critical point symmetry of the interacting boson
model.
\end{abstract}

\maketitle


\section{Introduction}

Recently, based on the Generalized Wick Theory\cite{wick}, a
nucleon-pair shell model(NPSM) was proposed\cite{npsm}, in which
nucleon pairs with various angular momenta are used as building
blocks. Since modern computers fail for the calculation in the full
shell model space for the medium weight and heavy nuclei, some
truncation scheme need to be used.

 The tremendous success of the
interacting boson model (IBM)~\cite{ibm-old}, suggests that $S$ and
$D$ pairs play a dominant role in the spectroscopy of low-lying
nuclear modes \cite{McGrory,Otsuka,Halse}. Therefore, one normally
truncates the full shell-model space  to the collective $S$-$D~$
pair subspace in the NPSM. The latter is called the $SD$-pair shell
model(SDPSM)\cite{npsm,sdpair,zhao-f}.

A crucial point in the SDPSM is the validity of the $S$-$D$ pair
truncation. In Ref.\cite{sm-ibm}, shell model foundations of the IBM
was reviewed  by Iachello, the results seem to indicate that the
$S$-$D$ pair truncation is a reasonable approximation to the full
shell model space. This problem was also studied  in \cite{12,13,14}
with the conclusion that the $S$-$D$ pair subspace works well in the
vibrational region, but in the deformed region, the inclusion of G
pairs is crucial. But Dr. Zhao's
work\cite{zhao-prc-62-014316,zhao-prc-66-041301} show that the
essential properties within the full shell model space survive in
the $S$-$D$ pair subspace. What's more, if a pure
quadrupole-quadrupole interaction and a reasonable collective
$S$-$D$-pair were used, the rotational behavior can be produced very
well within the $S$-$D$ pair subspace. The fact that the SDPSM can
describe the collectivity of low-lying states for nuclei around
A=130\cite{ba,ba2,zhao1,zhao2,zhao3,zhao07,referee} also imply that
the $S$-$D$ pair truncation is a good approximation to the full
shell model space.


Nuclei, as a mesoscopic system, have been found to possess
interesting geometric shapes, such as spherical (vibrational
($U(5)$), axially deformed ($SU(3)$), and $\gamma$-soft ($O(6)$),
which is usually described in terms of the Casten
triangle~\cite{pan4}. The search for signatures of transitions among
various shapes (phases) of atomic nuclei is an interesting subject
in nuclear structure theory. An understanding of such shape (phase)
transitions may provide insight into quantum phase transitions in
other mesoscopic systems~\cite{ibm-old}.

Theoretical study of shape phase transitions and critical point
symmetries in nuclei was mainly carried out
\cite{ibm-old,Ginicchio,1,2,3,prl-81-1191,E5,X5,4,prl-87-162501,
Warner2,5,rowe,rowe2,Cejnar,liu,pan,leviatan,zhangyuprcr} in the
interacting boson model for identical system(IBM-I)\cite{ibm-old}.
The investigations on nuclear shape phase transition and critical
point symmetry for identical nucleon system have also been carried
out with fermionic degrees of freedom in
\cite{Ginocchio-so7,Ginocchio-so8,rowe3,rowe4,liuzhang,Ginocchio}.
 Recently, investigations of the shape
phase transitions and critical point symmetries in nuclei were also
carried out\cite{2-fluit,2-prl-93-242502} in the proton-neutron
interaction boson model(IBM-II)\cite{ibm-old}.

Since the SDPSM is also built up from $SD$ pairs, it is expected
that the SDPSM can produce similar results to those of the IBM. Our
previous work show that the vibrational, rotational, and gamma-soft
spectra can be well reproduced\cite{prc05} similar to the $U(5)$,
$SU(3)$ and $SO(6)$ limiting spectra in the IBM. What's more, the
vibrational-rotational phase transition for identical system can
also be produced within the framework of the SDPSM with fermionic
degrees of freedom\cite{prc06}. Since nuclei are neutron-proton
coupled systems,  and a rich phase structure can be obtained in the
neutron-proton IBM, it is interesting to see if the phase
transitional patterns in the neutron-proton coupled system can be
produced in the SDPSM with fermionic degrees of freedom. This is the
aim of this paper.


\section{Model}

In the shell model description, the pairing and
quadrupole-quadrupole interactions are the most important
short-range and long-range correlations.
Considering that the Hamiltonian used to study the shape phase
transition in the IBM is mainly composed of the monopole pairing and
quadrupole-quadrupole interaction(e.g.,
Ref.\cite{2-fluit,2-prl-93-242502}),
a schematic Hamiltonian is adopted in the SDPSM, which is a
combination of the monopole pairing interaction and
quadrupole-quadrupole interaction with
\begin{eqnarray}
H_{X}&= &\sum_{\sigma=\pi,\nu} (-G_\sigma S_\sigma^\dagger S_\sigma
 -\kappa_\sigma Q_\sigma^{(2)} \cdot Q_\sigma ^{(2)}) -
 \kappa Q_\pi^{(2)} \cdot Q_\nu^{(2)},\\
{\cal S^\dagger} &=&\sum_a {\frac {\widehat a} 2}\left(C_a^\dagger
\times C_a^\dagger \right) \nonumber,\\
Q^{(2)} & = & \sqrt{16\pi /5}\sum_i r^{2}_{i} Y^{2}(\theta_i,\phi_i)
\nonumber
\end{eqnarray}
where $X$ in $H_{X}$ is denoted as $U(5)$, $SU(3)$, or $SO(6)$
corresponding vibrational, rotational, or gamma-soft limiting case
in the model, $G_\sigma$ and $\kappa_\sigma$ are the pairing  and
quadrupole-quadrupole interaction strength between
identical-nucleons, respectively. $\kappa$ is the
quadrupole-quadrupole interaction strength between proton and
neutrons. In this paper, we set $G_\pi=G_\nu$ and
$\kappa_\pi=\kappa_\nu$.

To study the phase transitional patterns, the Hamiltonian for
proton-neutron coupled system is written as
\begin{equation}
H=(1-\alpha)H_{U(5)} +\alpha H_X,
\end{equation}
where $0 \leq \alpha\leq 1$ is a control parameter, $H_X$ is taken
as $H_{SU(3)}$ when we study vibration-rotation transitional
patterns, and is taken as $H_{SO(6)}$ when we study vibration to
$\gamma$-soft transitional patterns.

The $E2$ transition operator adopted is
\begin{eqnarray}
T(E2) & = & e_\pi Q_\pi ^{(2)} +e_\nu Q_\nu^{(2)},
\end{eqnarray}
where $e_{\pi}(e_\nu)$ is the effective charge for proton(neutron).


The  collective $S$-pair is defined as
\begin{eqnarray}
S^\dagger & = & \sum_a y(aa0)(C_a^ \dagger \times C_a^\dagger)^0
\end{eqnarray}
In this paper, the $S$-pair structure coefficient, as an
approximation, is fixed to be $y(aa0)=\widehat a\sqrt{{\frac
{N}{\Omega _a -N}}}$, where $\Omega_a$ is defined as $\Omega _a
=a+1/2$ and $N$ is the number of pairs for like-nucleons. The
$D$-pair is obtained by using commutator
\begin{eqnarray}
 D^{\dagger} & =&
           \mbox{$\frac{1}{2}$} [Q^{(2)}, S^{\dagger} ] = \sum_{ab}^{} y(ab2)
           \left( C^{\dagger}_{a}\times C^{\dagger}_{b} \right)^{2}.
          \label{commutator}
\end{eqnarray}
After symmetrization, it is easy to obtain that
\begin{eqnarray}
y(ab2)=-{\frac 1 2}q(ab2)\left[{\frac {y(aa0)} {\hat a}}+{\frac
{y(bb0)} {\hat b}}\right].
\end{eqnarray}
The details of the model can be found in \cite{npsm,sdpair,zhao-f}

\section{Results}

To identify shape phase transitions and determine the corresponding
patterns, Iachello {\it et al} initialed a study on effective order
parameters, which should display different critical behaviors for
the phase transitions with different order. Specifically, the
quantities related with isomer shifts, defined as $v_2=(<0_2^+|\hat
n_d|0_2^+>-<0_1^+|\hat n_d|0_1^+>)/N$ and $v_2^\prime=(<2_1^+|\hat
n_d|2_1^+>-<0_1^+|\hat n_d|0_1^+>)/N$ , were proposed as
effective-order parameters in \cite{5}. Consequently, some other
quantities, such as the $B(E2)$ ratios $K_1=B(E2;4_1^+\rightarrow
2_1^+)/B(E2;2_1^+\rightarrow 0_1^+)$ and $K_2=B(E2;0_2^+\rightarrow
2_1^+)/B(E2;2_1^+\rightarrow 0_1^+)$ \cite{zhangyuprcr} as well as
the energy ratio $R_{60}=E_{6_1^+}/E_{0_2^+}$ were also suggested as
the effective order parameters to identify phase transitions and the
corresponding orders. Therefore, to study the shape phase transition
in the $SD$-pair fermion model space, $v_2$, $v_2^\prime$, in which
the d-boson number operator $\hat n_d$ is replaced by $D$-pair
number operator $\hat N_D$ in the SDPSM, $K_1$, $K_2$ and $R_{60}$
will be studied in this paper. Because the importance of
$R_{42}=E_{4_1^+}/E_{2_1^+}$ in determining the limiting cases and
shape phase transitions\cite{plb527}, $R_{42}$ is also presented.


\subsection{vibration-rotation transitional patterns}
We begin by considering the vibration-rotation phase transition. A
system with $N_\pi=N_\nu=3$ in $gds$ shell was studied.
By fitting $R_{42}\equiv E_{4_1^+}/E_{2_1^+}=2$ for vibrational
case, the parameters used to produce the vibrational spectra were
obtained, and presented in Table \ref{para}. The detailed discussion
about  the vibrational spectra can be found in \cite{prc05}.
In the SDPSM, the full shell model space was truncated to the
$SD$-pair subspace. The investigation on the validity of the $S$-$D$
pair truncation in \cite{12,13,14} show that the $S$-$D$ pair
truncation can not produce the rotational spectra. But Dr. Zhao's
work\cite{zhao-prc-62-014316,zhao-prc-66-041301} and our previous
work\cite{prc05} show that if  a pure quandrupole-quadrupole
interaction and a reasonable collective $S$-$D$ pair were used, the
rotational behaviors can be produced very well. It is found that
with $2\kappa_\pi=2\kappa_\nu=\kappa=0.2$MeV$/r_0^4$, the similar
results as the $SU(3)_\pi \times SU(3)_\nu$ limit of the IBM can be
produced, the typical energy ratios $E_{4_1^+}/E_{2_1^+}$ and $
E_{6_1^+}/E_{2_1^+}$ are 3.33 and 6.96, close to the IBM result 3.33
and 7. The detailed discussion can be found in
Ref.\cite{prc05,zhao-prc-62-014316}.

\begin{table}[!h]
\tabcolsep 0pt \caption{The parameters used to produce the
vibrational, rotational and $\gamma$-soft spectra. $G_\sigma$ is in
unit of MeV, $\kappa_\sigma$ and $\kappa$ are in unit of
MeV/$r_0^4$.} \vspace*{-12pt}
\begin{center} \def\temptablewidth{0.8\textwidth}
{\rule{\temptablewidth}{1pt}}
\begin{tabular*}{\temptablewidth}{@{\extracolsep{\fill}}ccccccc}
\label{para} &limit & $G_\pi$ & $G_\nu$) & $\kappa_\pi$ &
$\kappa_\nu$ & $\kappa$ \\\hline
vibration-rotation &vibration & 0.5 & 0.5 & 0 & 0 & 0.01    \\
&rotation  &  0 & 0 & 0.1 & 0.1 & 0.2 \\\hline
vibration-$\gamma$-soft& vibration & 0.5 & 0.5 & 0 & 0 & -0.01    \\
& $\gamma$-soft & 0.15 & 0.15 & 0 & 0 & -0.015
\\\hline
\end{tabular*}
 {\rule{\temptablewidth}{1pt}}
\end{center}
\end{table}

\begin{figure}
\includegraphics[width=10cm]{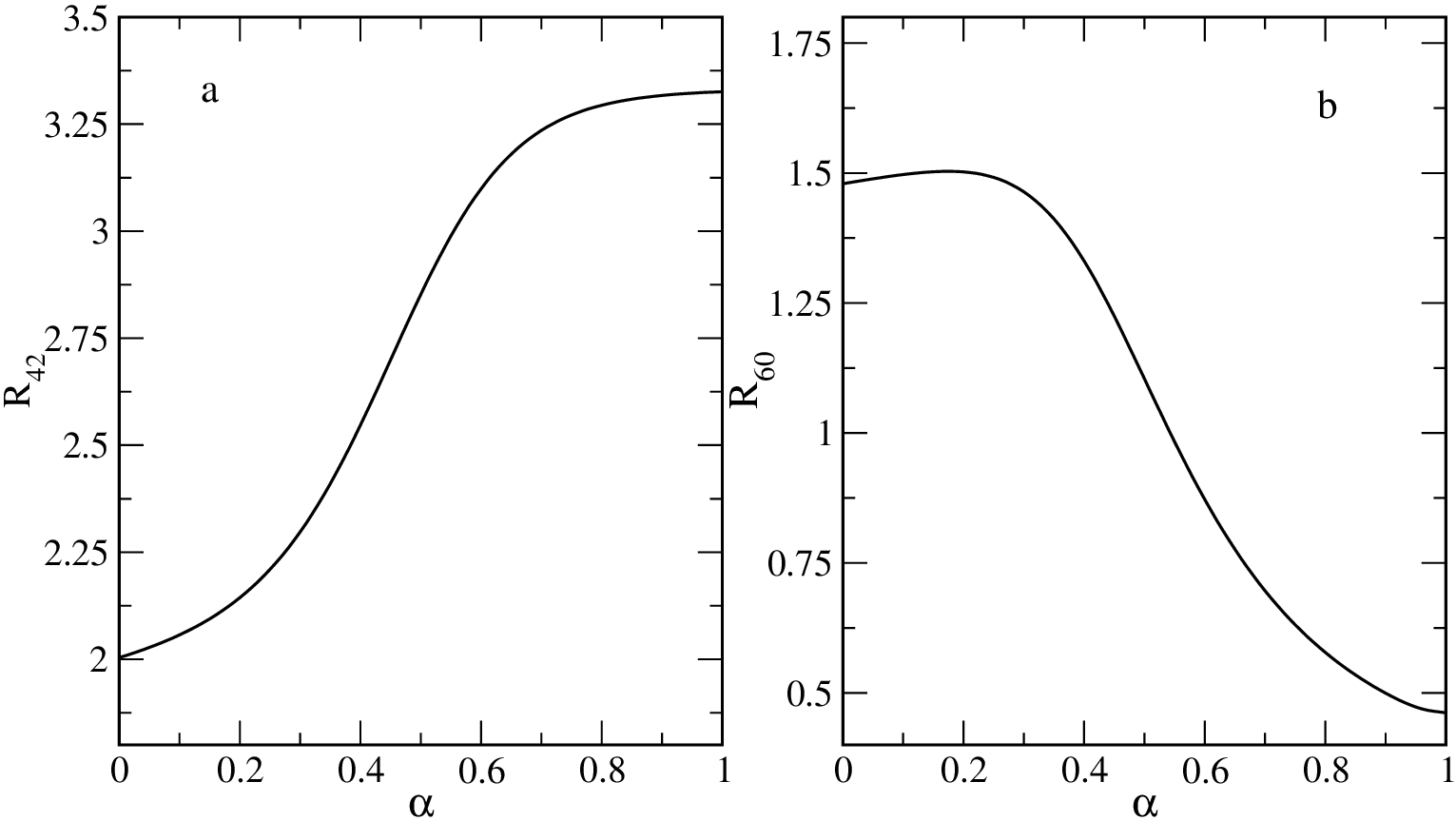}
\caption{Energy ratios $R_{42}$ and $R_{60}$ vs $\alpha$ for the
vibration-rotation transition.} \label{u5-u3-r42-r60}
\end{figure}

Energy ratios $R_{42}$ and $R_{60}$ against control parameter
$\alpha$ are shown in Fig.\ref{u5-u3-r42-r60}.
Fig.\ref{u5-u3-r42-r60}a shows that
 the energy ratio $R_{42}$ is 2(when $\alpha=0$) and 3.3
(when $\alpha=1$), typical values of vibrational and rotational
spectra in the IBM\cite{ibm-old}. It is also shown that the rapid
change occurs when $0.3 \leq \alpha \leq 0.6$, which indicates a
phase transition occurs in this region.

The energy ratio $R_{60}$ given in Fig.\ref{u5-u3-r42-r60}b shows
that similar behavior as that of the IBM for finite number of boson
$N_B$ is reproduced. It exhibits a modest peak followed by a sharp
decrease across the phase transition, a typical signature of the
1st-order quantum phase transition\cite{prl2008}.

\begin{figure}
\includegraphics[width=10cm]{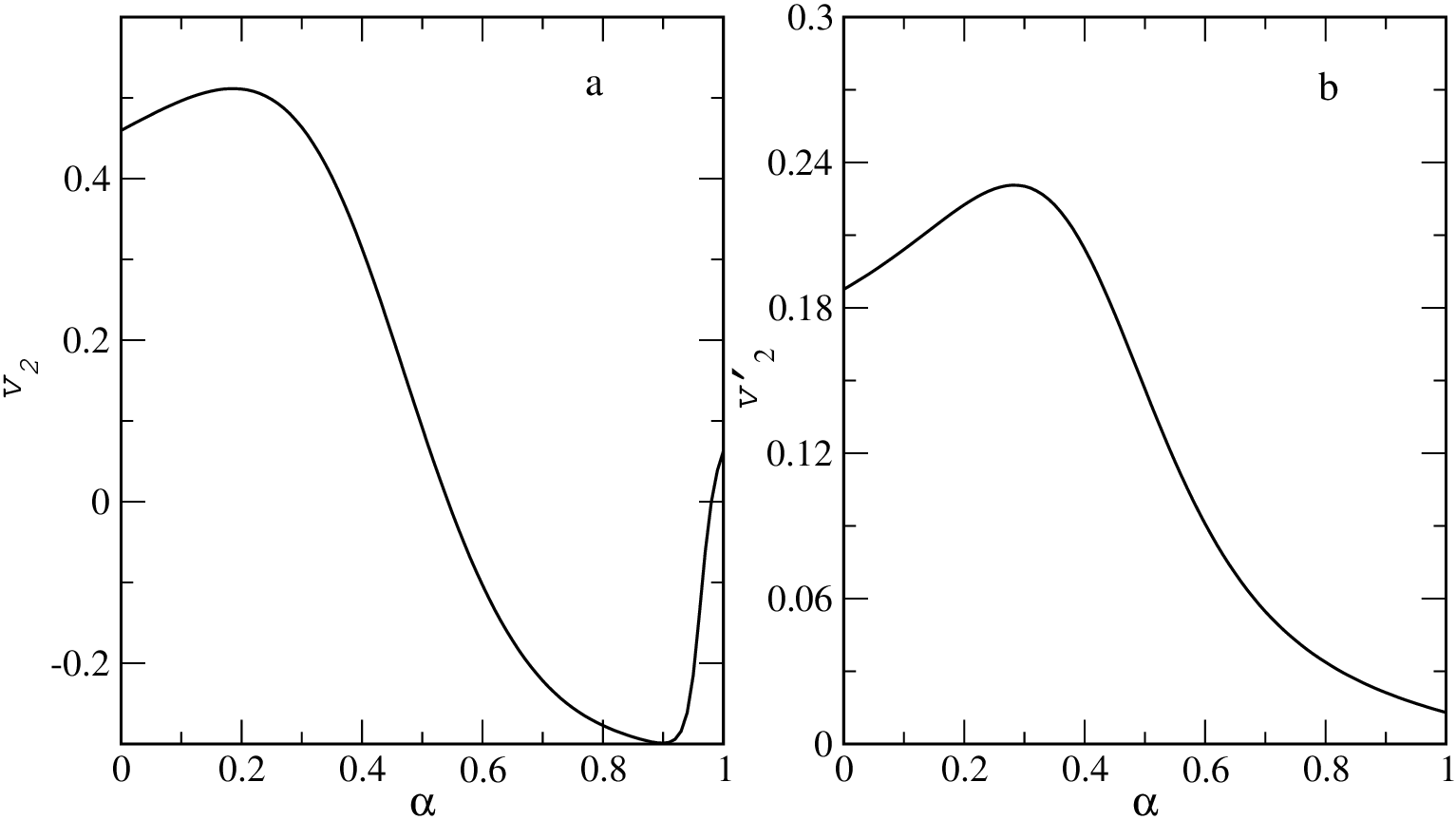}
\caption{$v_2$ and $v'_2$ vs $\alpha$ in the vibration-rotation
transition.} \label{u5-u3-v2-v2'}
\end{figure}

\begin{figure}
\includegraphics[width=10cm]{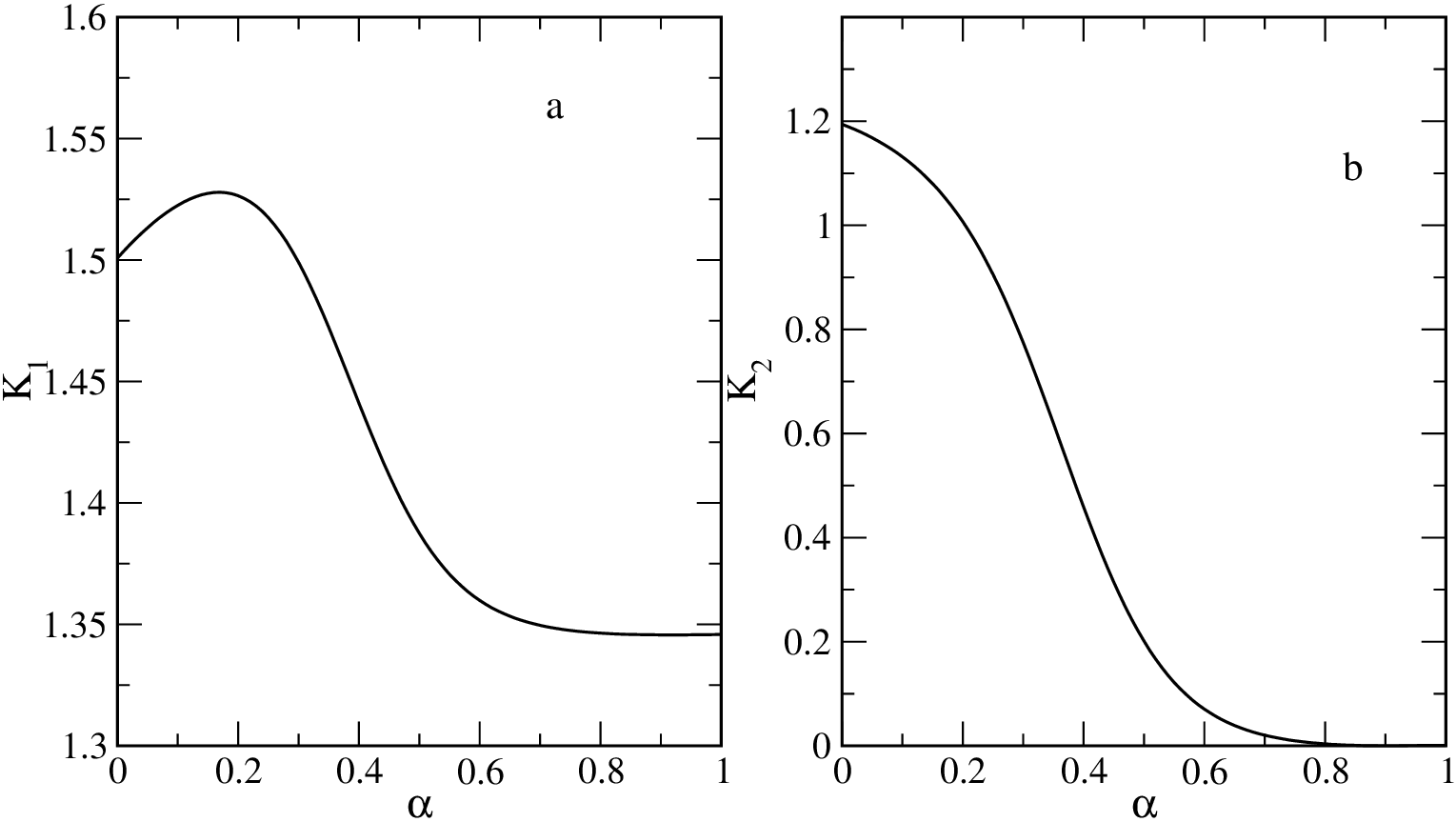}
\caption{B(E2) ratios vs $\alpha$ in the vibration-rotation
transition.} \label{u5-u3-k1-k2}
\end{figure}

The SDPSM results of $v_2$, $v_2^\prime$, $K_1$ and $K_2$ are given
in Fig.\ref{u5-u3-v2-v2'} and Fig.\ref{u5-u3-k1-k2}. The effective
charges were fixed with $e_\pi=3e_\nu=1.5e$. As argued in \cite{5},
$v_2$, $v_2^\prime$ should have wiggling behaviors in the region of
the critical point due to the switching of the two coexisting phases
for the first order phase transition, then the obvious wiggling
behaviors shown by $v_2$, $v_2^\prime$ in Fig.\ref{u5-u3-v2-v2'}
further confirm the transition is first order. The results of
$B(E2)$ ratio $K_1$ is consistent with those of other effective
quantities\cite{5,zhangyuprcr}. The critical behavior of $K_2$ seems
to deviate from the character of the first order phase transition.

In the IBM, the critical point symmetry\cite{X5} between $U(5)$ and
$SU(3)$ is $X(5)$. Since the shape phase transition between
vibrational and rotational limit can be reproduced in the SDPSM, it
is interesting to see if the properties of the $X(5)$-like symmetry
also occurs within the SDPSM. We found that there is indeed a
signature  with $\alpha=0.54$ in the SDPSM similar to that of the
$X(5)$ in the IBM. A few typical values are given in Table \ref{X5},
from which one can see that typical feature of the $X(5)$ symmetry
stated in Ref.\cite{prc68,prl2008} indeed occurs in the SDPSM. For
example, $R_{42}$, $R_{60}$ and $E_{0_2^+}/E_{2_1^+}$ is 2.91, 1.05
and 5.32 in the SDPSM calculation, close to the IBM results 2.91,
1.0 and 5.67, respectively.

\begin{table}[!h]
\tabcolsep 0pt \caption{Energy and B(E2) ratios at vibrational,
rotational limit, and $X(5)$-like critical point calculated in the
SDPSM.} \vspace*{-12pt}
\begin{center} \def\temptablewidth{0.8\textwidth}
{\rule{\temptablewidth}{1pt}}
\begin{tabular*}{\temptablewidth}{@{\extracolsep{\fill}}cccccc}
\label{X5}
   limit & $\frac{E_{4_1^+}}{E_{2_1^+}}$ & $\frac{E_{6_1^+}}{E_{2_1^+}}
   $& $\frac{E_{6_1^+}}{E_{0_2^+}}$& $\frac{4_1^+\rightarrow 2_1^+}{2_1^+\rightarrow 0_1^+}$
   &$\frac{6_1^+\rightarrow 4_1^+}{2_1^+\rightarrow 0_1^+}$  \\\hline
    vibrational limit & 1.99    & 2.97  &1.47&  1.49  & 1.48   \\
   $X(5)$-like point  &  2.91   & 5.60  &1.05& 1.38   & 1.38 \\
   rotational limit &  3.33   & 6.96  &0.46&1.34   & 1.32  \\\hline
  & $\frac{E_{0_2^+}}{E_{2_1^+}}$ & $\frac{E_{2^+}-E_{0_2^+}}{E_{2_1^+}}$ & $\frac{E_{4^+}-E_{0_2^+}}{E_{2_1^+}}$ &
$\frac{2^+\rightarrow 0_2^+}{2_1^+\rightarrow 0_1^+}$ & $\frac{4^+\rightarrow 2^+}{2_1^+\rightarrow 0_1^+}$ \\\hline
 $X(5)$-like point ($0_2^+$ band) & 5.32 & 2.30 & 5.33 & 0.37  & 0.43 \\\hline
\end{tabular*}
 {\rule{\temptablewidth}{1pt}}
\end{center}
\end{table}

\subsection{vibration-$\gamma$-soft transitional patterns}

The investigation on vibration-$\gamma$-soft shape phase transition
in the IBM has been studied in \cite{prl85-2000},  the corresponding
quantum phase transition was suggested to be of the 2nd-order.
Recently, similar phase transition within the fermion model for
identical nucleon system has also been
performed\cite{Ginocchio,liuzhang}.

From the periodic chart, one can deduce that nuclei that display an
SO(6) spectrum lie close to the end of the shell, at least in the
neutron sector. Therefore, to explore whether the transitional
patterns between vibration and $\gamma$-soft spectrum can be
realized in the SDPSM, we considered a system with $N_\pi=\tilde
N_\nu=3$ in the $gds$ shell. Namely, neutron pairs in this case were
treated as three neutron-hole pairs and a negative $\kappa$ was used
as in \cite{prc05}.
By fitting $R_{42}=2$ and $2.5$ for vibrational and $\gamma$-soft
limiting cases, the parameters were fixed, and the results are
listed in Table \ref{para}. The detailed discussion about the two
limiting cases in the SDPSM can be found in \cite{prc05}.

\begin{figure}
\includegraphics[width=10cm]{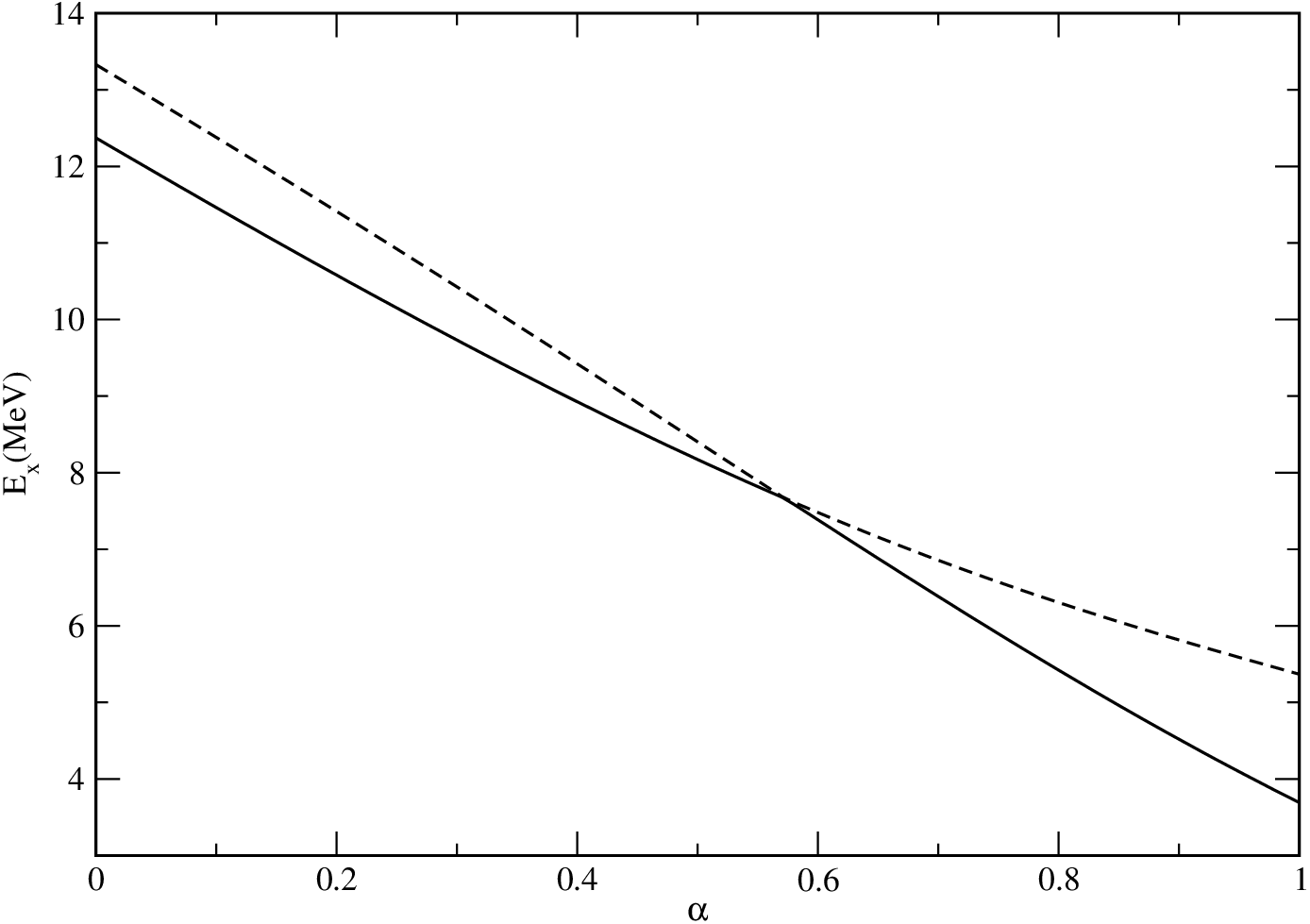}
\caption{Energy levels of $0_2^+$ and $0_3^+$ states vs $\alpha$ in
the vibration-$\gamma$-soft transitional region.} \label{02-03}
\end{figure}

The IBM calculation show that the level crossing-repulsion behavior
of $0_2^+$ and $0_3^+$ occurs\cite{pan-ijmpe} in the critical region
of the $U(5)$-$SO(6)$ transition. The SDPSM results of $0_2^+$ and
$0_3^+$ states, given in Fig.\ref{02-03}, show the similar behavior
of level crossing-repulsion when $\alpha=0.58$. Therefore, to see
the behavior of effective order parameters against the control
parameters clearly, the quantities related to $0_2^+$ state were
also calculated for the $0_3^+$ state.

\begin{figure}
\includegraphics[width=10cm]{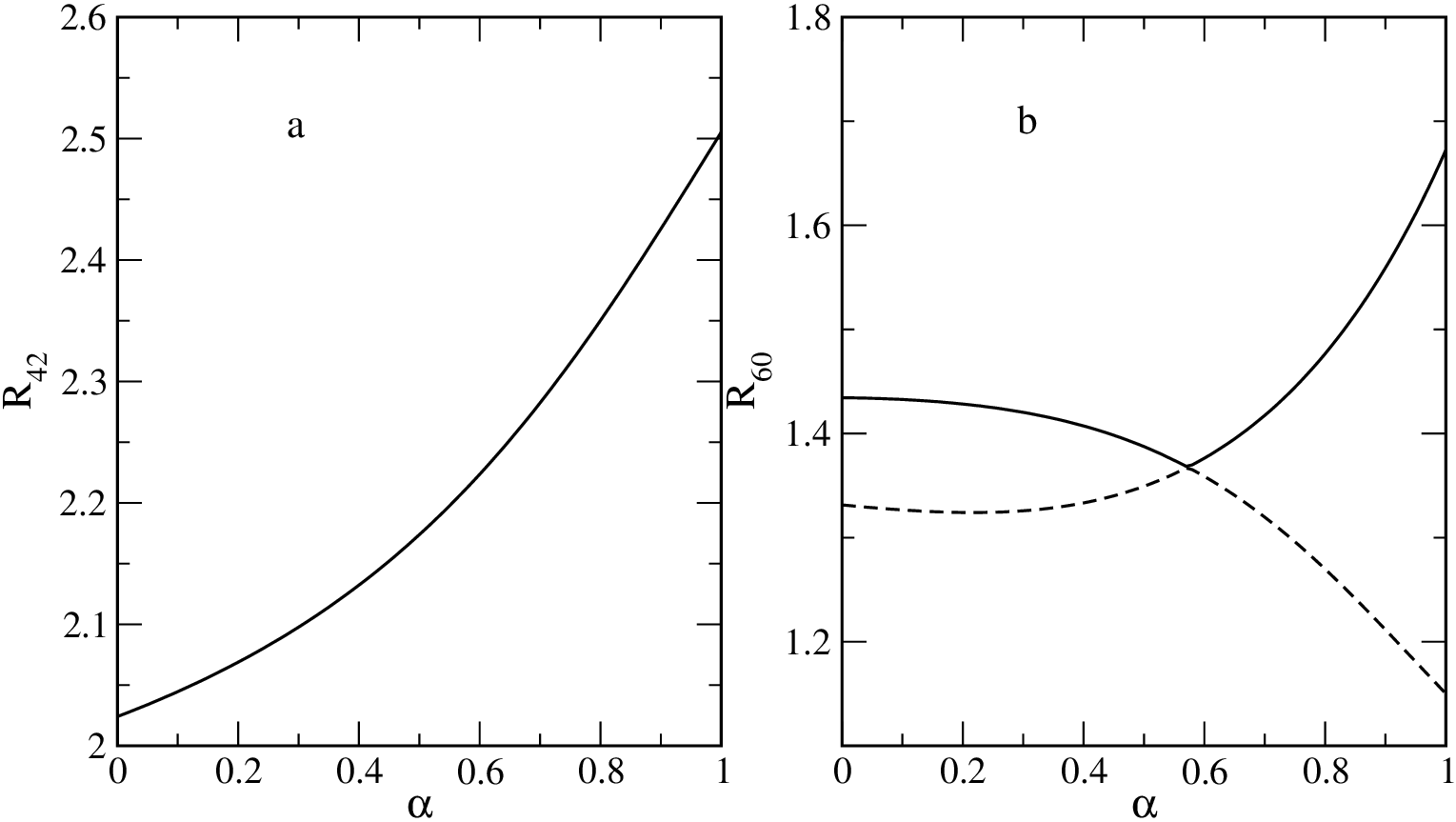}
\caption{Energy ratios $R_{42}$ and $R_{60}$ vs $\alpha$ in the
vibration-$\gamma$-soft transitional region.} \label{u5-o6-r42-r60}
\end{figure}

\begin{figure}
\includegraphics[width=10cm]{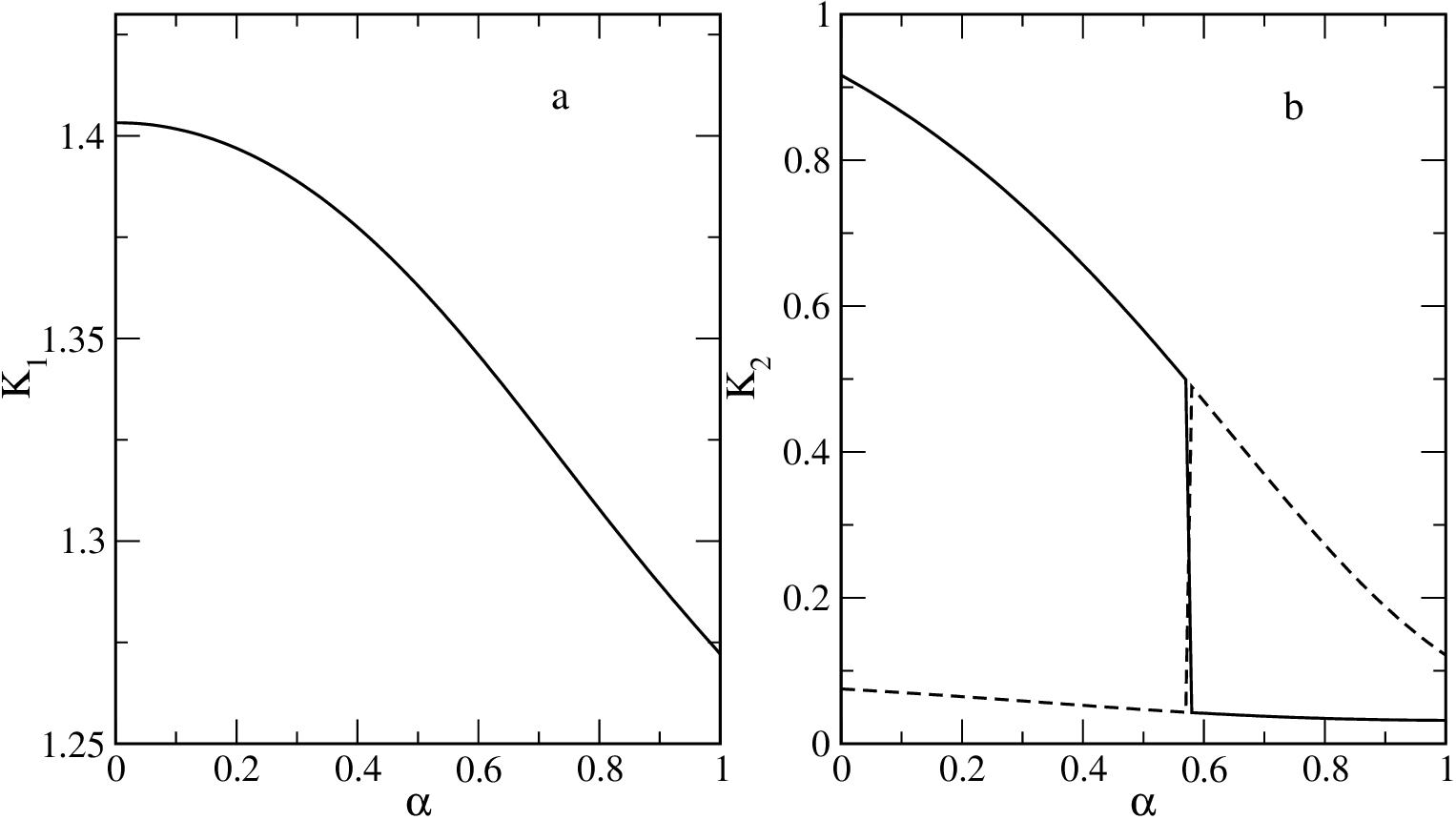}
\caption{$B(E2)$ ratios vs $\alpha$ in the vibration-$\gamma$-soft
transitional region.} \label{u5-o6-k1-k2}
\end{figure}

\begin{figure}
\includegraphics[width=10cm]{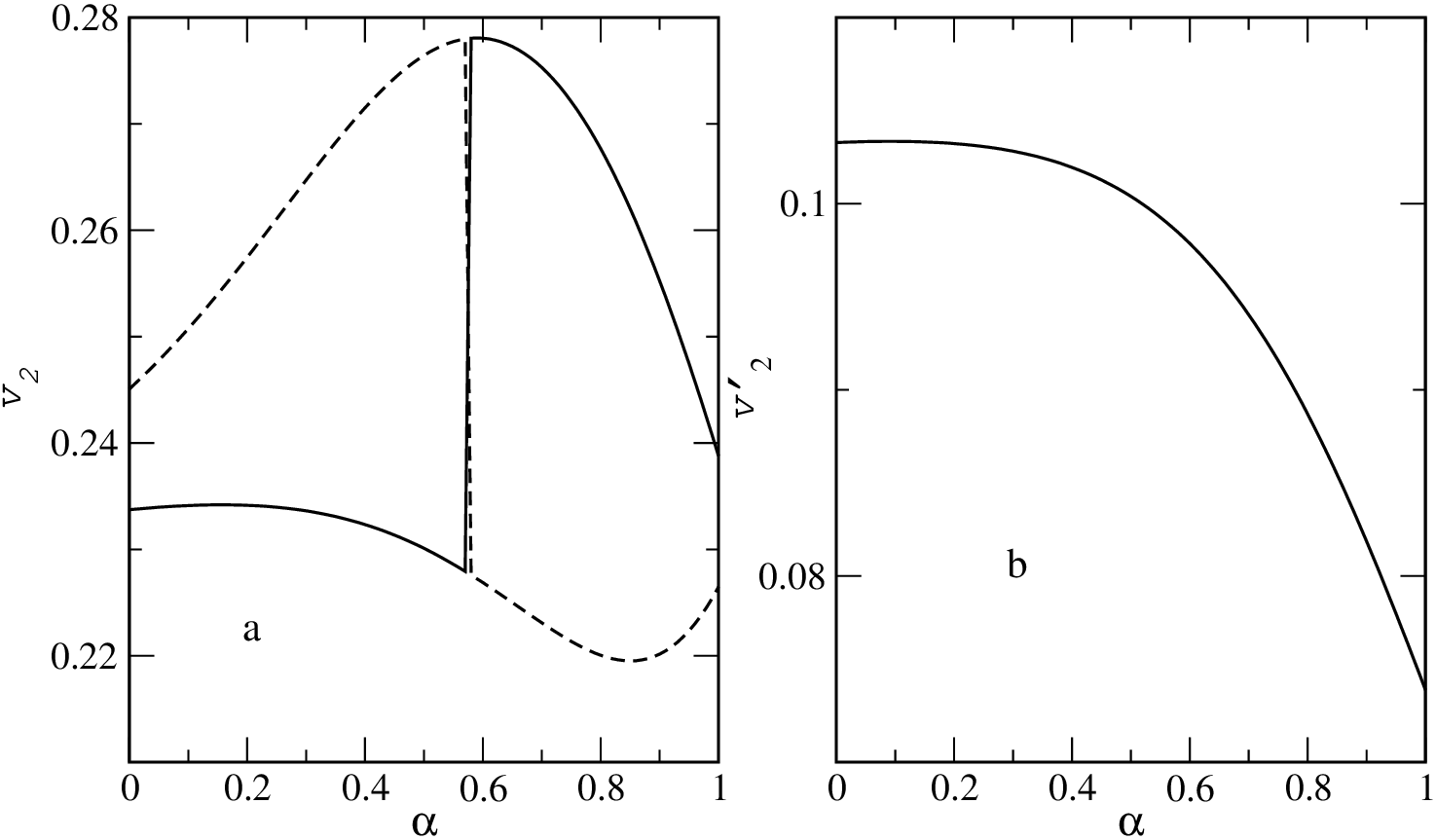}
\caption{$v_2$ and $v'_2$ vs $\alpha$ in the vibration-$\gamma$-soft
transitional region.} \label{u5-o6-v2-v2'}
\end{figure}

The results for $R_{42}$($R_{60})$, $K_1$($K_2)$ and
$v_2$($v_2^\prime)$ are given in Fig.\ref{u5-o6-r42-r60},
 Fig.\ref{u5-o6-k1-k2}  and Fig.\ref{u5-o6-v2-v2'},
respectively. The effective charges were fixed as
$e_\pi=-3e_\nu=1.5e$  since the neutron pairs were treated as holes.

Fig.\ref{u5-o6-r42-r60}a shows that the typical ratios,
$R_{42}=2$(when $\alpha=0)$ and $2.47$ (when $\alpha=1)$, of
vibration and $\gamma$-soft spectra were produced. Interestingly, we
found that in comparison with that of the rotation-vibration
transitional results, $R_{42}$ in the vibration-$\gamma$-soft
transitional region increases with $\alpha$ smoothly.

From Fig.\ref{u5-o6-k1-k2} one can see that as the IBM
results\cite{zhangyuprcr} and $R_{42}$ given in
Fig.\ref{u5-o6-r42-r60}a, the wiggling behavior in $K_{1}$ is
smoothed out in the vibration-$\gamma$-soft transition. One can also
see that because the structure of $0_2^+$ and $0_3^+$ exchange at
$\alpha \backsim 0.58$, the amplitudes of $B(E2;0_2^+\rightarrow
2_1^+)$ and $B(E2;0_3^+\rightarrow 2_1^+)$ also exchange at this
point.

In \cite{prl2008}, the experimental data of
 Xe and Ba isotopes were analyzed, in which for smaller neutron
 numbers, $^{134,136}$Ba and $^{128}$Xe, the $0_3^+$ state was taken
 in the $R_{60}$ if its $B(E2)$
 decay was consistent with $\sigma=N-2$.
It was also shown that\cite{pan-ijmpe}
 $B(E2;0_2^+\rightarrow
2_1^+)/B(E2;2_1^+\rightarrow 0_1^+)=0.07$ for $^{196}$Pt, while it
is $0.81$ for $^{198}$Pt. By considering these results, the $0_3^+$
state were taken in the $R_{60}$, $K_2$ and $v_2^\prime$ when
$\alpha>0.58$. In comparison with those in the vibration-rotation
transition, Fig.\ref{u5-o6-r42-r60}b and Fig.\ref{u5-o6-k1-k2}a
 show that  $R_{60}$, $K_1$, and $K_2$  change smoothly with
 $\alpha$, which are the
 typical features of the 2nd-order phase transition\cite{prl2008,zhangyuprcr}.

Fig.\ref{u5-o6-v2-v2'} shows that as predicted in the IBM and shell
model calculation for identical system, the vibration-$\gamma$-soft
phase transition takes place and it is the second order phase
transition, for which $v_2$ and $v_2^\prime$ change smoothly with
$\alpha$, the wiggling behavior changing sign in the region of the
critical point are smoothed out.

In the $U(5)$-$SO(6)$ transitional region in the IBM, $E(5)$ is the
critical point symmetry\cite{prl85-2000,Leviatan-Ginocchio}. It is
interesting to see whether the signature of the $E(5)$-like symmetry
can be realized in the SDPSM for proton-neutron coupled system. We
found that $E_{4_1^+}/E_{2_1^+}=2.19$ when $\alpha=0.54$
corresponding to the typical value of $E(5)$ symmetry in the IBM.
Other typical results are listed in Table \ref{E5}, in which the IBM
results for $N=5$ are also given\cite{prl85-2000}. It is seen that
except for $E_{0_2^+}/E_{2_1^+}=2.59$, which is smaller than that of
the IBM result for $N=5$, the properties of $E(5)$ symmetry in the
IBM indeed occurs in the SDPSM.

\begin{table}[!h]
\tabcolsep 0pt \caption{The SDPSM results for $E(5)$-like symmetry.
The corresponding results with $N=5$ in the IBM are also
given\cite{prl85-2000}.} \vspace*{-12pt}
\begin{center} \def\temptablewidth{0.8\textwidth}
{\rule{\temptablewidth}{1pt}}
\begin{tabular*}{\temptablewidth}{@{\extracolsep{\fill}}ccccccccc}
\label{E5}
   limit & $E_{4_1^+}/E_{2_1^+}$ & $E_{0_2^+}/E_{0_3^+}$ & $E_{0_2^+}/E_{2_1^+}$ \\\hline
    SDPSM & 2.19  & 0.99  &  2.59  \\
     IBM  & 2.19  & 1.04  &  3.68   \\\hline

    & $\frac {4_1^+\rightarrow 2_1^+}{2_1^+\rightarrow 0_1^+}$
    &$\frac{2_2^+\rightarrow 2_1^+}{2_1^+\rightarrow 0_1^+}$
    & $\frac{0_2^+\rightarrow 2_1^+}{2_1^+\rightarrow 0_1^+}$\\\hline
SDPSM &  1.36 & 1.29 & 0.53\\
IBM   & 1.38  & 1.39 & 0.51 \\\hline
    & $\frac{0_2^+\rightarrow 2_2^+}{0_2^+\rightarrow 2_1^+}$ &
     $\frac{0_3^+\rightarrow 2_1^+}{0_3^+\rightarrow 2_2^+}$ & \\\hline
SDPSM &  0.06 & 0.03  &  \\
IBM  &  0 & 0 & \\\hline
\end{tabular*}
 {\rule{\temptablewidth}{1pt}}
\end{center}
\end{table}

\section{Summary}

In summary, the shape phase transition patterns for proton-neutron
coupled system were studied within the framework of the SD-pair
shell model. The results show that patterns of vibration-rotation
and vibration-$\gamma$-soft shape phase transitions are indeed
similar to the corresponding results obtained from the IBM
previously. The signatures of the critical point symmetry in the
SD-pair shell model are also close to those shown in the IBM. The
procedure may be extended to study quantum phase transitions in
other fermion systems.

\vskip 0.5cm This work was supported in part by the Natural Science
Foundation of China (10675063; 10775064), the U.S. National Science
Foundation (0140300; 0500291), the Education Department of Liaoning
Province (20060464), and the LSU--LNNU joint research program
(LSU-9961).

\end{document}